\documentclass{webofc}
\usepackage[varg]{txfonts} 
\usepackage{subfigure}
\usepackage{float}
\newcommand{\indep}{\perp \!\!\! \perp}
\begin{document}

\title{Relativistic Dissipative Magnetohydrodynamics from the Boltzmann equation for 2-particle species gas}

\author{\firstname{Khwahish} \lastname{Kushwah}\inst{1}\fnsep\thanks{\email{khwahish_kushwah@id.uff.br}} \and
        \firstname{Gabriel} \lastname{S.~Denicol}\inst{1}\fnsep\thanks{\email{gsdenicol@id.uff.br}}
        }

\institute{Instituto de F\'isica, Universidade Federal Fluminense (UFF), Niter\'oi,
24210-346, RJ, Brazil
          }

\abstract{
We derive the equations of motion of relativistic magnetohydrodynamics from the Boltzmann equation using the method of moments. We consider a locally electrically neutral system composed of two particle species with opposite charges, with vanishing dipole moment or spin, so that the fluid has vanishing magnetization and polarization. We find that the dynamics of this fluid changes dramatically in the presence of a magnetic field. The shear stress tensor no longer adheres to a single differential equation; instead, it splits into three non-degenerate components, each evolving according to distinct dynamical equations. Exploring these equations in a Bjorken flow scenario, we find that for large magnetic fields, our theory predicts oscillatory behavior beyond the scope of an Israel-Stewart-like theory.
}
\maketitle
\section{Introduction}
\label{intro}
Relativistic magnetohydrodynamics (RMHD) provides a theoretical framework for understanding the dynamics of relativistic fluids in the presence of strong magnetic fields. This framework is crucial in various astrophysical scenarios, including high-energy heavy-ion collisions \cite{Armas_2022}. In the early stages of these collisions, intense magnetic fields are generated, reaching peaks of $10^{19}$ Gauss (RHIC) and $10^{20}$ Gauss (LHC) \cite{Hattori_2022}. While previous studies have primarily focused on single-particle fluids \cite{Denicol_2018,Denicol_resistive,nonResistive_chapman,Resistive}, we explore a very simple yet more realistic scenario involving a two-species fluid of massless particles with opposite charges. We show that the derived equations of motion differ considerably to the traditional fluid-dynamical equations for the shear stress tensor. Instead, we find that different components of the shear stress tensor, decomposed with respect to the direction of the magnetic field, satisfy distinct evolution equations. Notably, under a moderately strong magnetic field, the shear stress tensor exhibits oscillatory dynamics, deviating from conventional hydrodynamics and magnetohydrodynamics approaches, challenging standard Israel-Stewart-type theories and calling for a more nuanced understanding of the dynamics in electrically charged fluids.

\section{Equations of motion}
\label{Boltzmann equation}
We consider a relativistic locally electrically neutral fluid composed of two types of massless classical particles with opposite electric charges and vanishing dipole moment or spin, so that the fluid has vanishing magnetization and polarization. We also assume vanishing electric charge chemical potential, i.e., $\mu$ = 0. For this system the Boltzmann equation reads,
\begin{equation}
\setlength\abovedisplayskip{3pt}
\setlength\belowdisplayskip{3pt}
        k^{\mu} \partial_\mu f^{\pm}_{k} + q^{\pm} k_{\nu} F^{\mu \nu} \frac{\partial}{\partial k^{\mu}}f_k^{\pm}  = C[f^{\pm},f^\mp],
\end{equation}
where we consider only elastic collisions, 
\small\begin{equation}
    C[f^\mp,f^\pm] \equiv \frac{1}{2} \int dK^{'} dP dP^{'} W^{\mp\mp}_{KK^{'} \leftrightarrow PP^{'}} \left( f_p^\mp f_{p^{'}}^\mp - f_k^\mp f_{k^{'}}^\mp \right) + \int dK^{'} dP dP^{'} W^{\mp\pm}_{KK^{'} \leftrightarrow PP^{'}} \left( f_p^\mp f_{p^{'}}^\pm - f_k^\mp f_{k^{'}}^\pm \right).
\end{equation}\normalsize
 The index $\pm$ designates particle species, where '+' represents particles with positive charges and '-' represents particles with negative charges. Above, $W^{\mp\mp}_{KK^{'} \leftrightarrow PP^{'}}$ is the transition rate and $f^\pm$ is the single particle distribution function for $\pm$ particle species, respectively. Further, $F^{\mu\nu}$ is the Faraday tensor which is decomposed as $\displaystyle{
     F^ {\mu\nu} \equiv E^\mu u^\nu - E^\nu u^\mu +\epsilon^{\mu\nu\alpha\beta}u_\alpha B_\beta.}$
 Here we defined electric field four-vector, $E^\mu$ $\equiv$ $F^{\mu\nu}\ u_\nu$ and magnetic field four-vector, $B^\mu$ $\equiv$ $\epsilon^{\mu\nu\alpha\beta} F_{\alpha\beta} \ u_\nu/2$.

\subsection{Exact equations of motion}
\label{sub:exact eom}
For a fluid consisting of two particle species, we directly calculate the time derivative of the total shear stress tensor, $\pi^{\mu\nu}=\pi_+^{\mu\nu}+ \pi_-^{\mu\nu}$, and the relative shear stress tensor of the whole system, $\delta\pi^{\mu\nu}\equiv \pi_+^{\mu\nu}- \pi_-^{\mu\nu}$ where $\pi^{\mu\nu}_\pm$ denotes the individual shear stress tensor for each particle species. We have used the 14-moment approximation, together with the assumption of massless particles to calculate the collision term \cite{DNMR}. The resultant equations of motion  are
\begin{align}
\label{totalshear}
\Delta _{\alpha \beta }^{\mu \nu }\dot{\pi}^{\alpha \beta
}+\Sigma\pi^{\mu \nu }+\frac{2|q|B}{5T}b^{\lambda
\langle \mu}\delta\pi_{\lambda }^{\nu \rangle
}&=\frac{8}{15}\epsilon\sigma ^{\mu \nu }-\frac{4}{3}\pi^{\mu
\nu }\theta -\frac{10}{7}\sigma ^{\lambda \langle \mu }\pi
_{\lambda }^{ \nu \rangle} -2\omega^{\lambda\langle\nu}\pi^{\mu \rangle}_{\lambda},\\
\label{deltashear}
\Delta _{\alpha \beta }^{\mu \nu }\delta\dot{\pi}^{\alpha \beta
}+\Sigma^{'}\delta\pi^{\mu \nu }+\frac{2|q|B}{5T}b^{\lambda
\langle \mu}\pi_{\lambda }^{\nu \rangle} &= -\frac{4}{3}\delta\pi^{\mu
\nu }\theta -\frac{10}{7}\sigma ^{\lambda \langle \mu}\delta\pi
_{\lambda }^{\nu\rangle} -2\omega^{\lambda\langle\nu}\delta\pi^{\mu \rangle}_{\lambda}.\hspace{1.2 cm}
\end{align}
Above, we defined energy density, $\displaystyle{ \epsilon \equiv u_\mu u_\nu T^{\mu\nu}}$, the shear tensor, $\displaystyle{\sigma^{\mu\nu} \equiv \nabla^{\langle\mu} \ u^{\nu\rangle}}$, the expansion scalar, $\displaystyle{\theta \equiv \nabla_\mu u^\mu}$ and the vorticity tensor, $\displaystyle{\omega^{\mu\nu} = (\nabla^{\mu}u^\nu - \nabla^\nu u^\mu)/2}$, with $u^\mu$ being the fluid four-velocity and $\nabla_\mu = \Delta^\nu_\mu\partial_\nu$ the spatial projected gradient. Further, we have defined two transport coefficients:
$\displaystyle{
\Sigma =3n_0(\sigma_T^{+-}+\sigma_T)/5 \ \text{and} \ \Sigma^{'}=n_0(5\sigma_T^{+-}+3\sigma_T)/5,
}$ utilized later where $\sigma_T^{+-}$ denotes the cross section for different species particle collision and $n_0$ is particle density of individual particle species (since $\mu$ = 0). We further assume $\sigma^{++}_T = \sigma^{--}_T \equiv \sigma_T$, representing the total cross section, which is also assumed to be constant for simplifying the derivation of fluid dynamical equations.
We have also employed the notation, $\displaystyle{A^{\langle\mu\nu\rangle} \equiv \Delta^{\mu\nu}_{\alpha\beta}A^{\alpha\beta}}$, with $\displaystyle{\Delta^{\mu\nu}_{\alpha\beta}  \equiv \left(\Delta^{\mu}_\alpha \Delta^{\nu}_\beta + \Delta^{\nu}_\alpha\Delta^{\mu}_\beta - 2/3\Delta^{\mu\nu}\Delta_{\alpha\beta}\right)/2}$, $\displaystyle{\Delta^{\mu\nu} = g^{\mu\nu} - u^{\mu}u^{\nu}}$ and $\displaystyle{ g^{\mu\nu} = diag\ (1,-1,-1,-1)}$. Further, $b^{\mu\nu}$ = $\epsilon^{\mu\nu\alpha\beta} u_\alpha b_\beta$ is a dimensionless antisymmetic tensor where $b^\mu \equiv$ $B^\mu / B$ specifies the direction of magnetic field.

\subsection{New Projections and definitions}
\label{sub:newprojections}
Now focusing on the shear-stress tensor, we proceed to decompose it with respect to the direction of the magnetic field, $b^\mu$, in a complete, normalized, and orthogonal basis $(u^{\mu },\ b^{\mu },\ \ell^{\mu}_\pm)$:
\begin{equation}
\begin{split}
\label{projmag}
\pi ^{\mu \nu }=\pi _{\parallel }\left( b^{\mu }b^{\nu } +\frac{1}{2}\Xi
^{\mu \nu }\right) +2\pi_{\perp}^- b^{\left( \mu \right.}\ell_-^{\left. \nu
\right)} + 2\pi_{\perp}^+ b^{\left( \mu \right.}\ell_+^{\left. \nu
\right)}+\pi_{\indep}^+ \ell_+^{\mu}\ell_+^\nu +\pi_{\indep}^- \ell_-^{\mu}\ell_-^\nu,
\end{split}
\end{equation}
where we defined the projection operator, $\displaystyle{\Xi ^{\mu \nu } \equiv g^{\mu \nu }-u^{\mu }u^{\nu }+b^{\mu }b^{\nu } = \Delta^{\mu\nu} +b^{\mu}b^{\nu}}$, onto the subspace orthogonal to $u^{\mu }$ and $b^{\mu }$, and $\displaystyle{g^{\mu\nu} = u^\mu u^\nu -b^\mu b^\nu - \ell^\mu_+ \ell^\nu_- - \ell_-^\mu\ell_+^\nu}$. Here, $\ell^{\mu}_\pm$ is the plane orthogonal to the magnetic field in the local rest frame of the fluid such that
\begin{equation}
u^{\mu } =\left( 1,0,0,0\right) ,  \qquad
\ell^{\mu}_\pm =\left( 0,1,\pm i,0\right)/\sqrt{2} ,  \qquad
b^{\mu } =\left( 0,0,0,1\right),
\end{equation}

It is important to note that now the '$\pm$' index no longer denotes the particles species but rather our convention for the projections into the subspace orthogonal to the magnetic field and fluid 4-velocity.

Using the definitions from Eq.\eqref{projmag}, we take required projections of Eqs.\eqref{totalshear} and \eqref{deltashear} to obtain different components of shear stress tensor and rewrite them in new basis to get equations of motion for longitudinal ($\pi_\parallel$), semi-transverse ($\pi_{\perp} \equiv \pi^{-}_{\perp} + \ \pi^{+}_{\perp}$) and transverse component ($\pi_{\indep} \equiv \pi^{-}_{\indep} + \ \pi^{+}_{\indep}$) of shear stress tensor. In the following section, we shall investigate these equations in a simple dynamical model, Bjorken flow \cite{Bjorken}, to observe the behaviour of different components of shear stress tensor for moderately large values of magnetic field.

\section{Bjorken Flow}
\label{sec:Bjorken flow}
We have established that the shear stress tensor comprises three distinct components, and as we shall see, each component evolves in a unique manner. To examine these components in Bjorken flow, we consider the magnetic field, $b^\mu$ in transverse direction. The shear tensor in this coordinate is expressed as following where only its spatial diagonal elements survive \cite{Denicol_Rischke}, 
\begin{equation}
\setlength\abovedisplayskip{5pt}\setlength\belowdisplayskip{5pt}
    \sigma_{\mu\nu} = \mathrm{diag} \left(0,\frac{1}{3\tau},\frac{1}{3\tau},-\frac{2\tau}{3} \right),
\end{equation}
with the expansion rate, $\displaystyle{\theta = 1/\tau}$. The shear tensor can also be decomposed according to \eqref{projmag}, leading to
\begin{subequations}
\setlength\abovedisplayskip{-3pt}
\setlength\belowdisplayskip{5pt}
\begin{align}
    \sigma_{\parallel}&= b_{\mu}b_{\nu}\sigma^{\mu\nu} = \frac{1}{3\tau},\\
    \sigma^{\pm}_{\perp} & = \ell ^{\mp}_{\mu} b_{\nu}\sigma^{\mu\nu} = 0,\\
    \sigma^{\pm}_{\indep} & = \ell^{\pm}_{\mu}\ell^{\pm}_{\nu} \sigma^{\mu\nu}_{\indep} = \ell^{\pm}_{\mu}\ell^{\pm}_{\nu} \sigma^{\mu\nu} = \frac{1}{2\tau}.
    \end{align}
    \end{subequations}
Since, $\sigma^{\pm}_{\perp} = 0$, the semi-transverse component has no impact on the evolution of the other two components. Thus, our focus is solely on the dynamics of the longitudinal and transverse components. The equation of state, pertaining to two types of particles, each with three quarks and two spins, is defined as:
\begin{equation}
\setlength\abovedisplayskip{5pt}\setlength\belowdisplayskip{5pt}
    \epsilon  = \frac{3\times 2\times 2\times  3}{\pi^2}T^4.
\end{equation}
Furthermore, from Maxwell's equations, we have that (see also \cite{Roy_2015}),
\begin{equation}\label{Bfield}
\setlength\abovedisplayskip{5pt}\setlength\belowdisplayskip{5pt}
    \dot{B} + B\theta = 0\quad \Longrightarrow \quad B \sim \left(\frac{\tau_0}{\tau}\right).
\end{equation}
The second-order equations of motion takes following form in Bjorken flow:
\setlength\abovedisplayskip{5pt}\setlength\belowdisplayskip{5pt}
\begin{align}
      \frac{d \epsilon}{d\tau}  & 
      = \frac{\pi_\parallel}{2\tau} + \frac{\pi_{\indep}}{2\tau}-\frac{4\epsilon}{3\tau},\label{energy}\\
      \frac{d}{d\tau}\left(\frac{\pi_\parallel}{\epsilon}\right) + \Sigma\frac{\pi_\parallel}{\epsilon}& = \frac{8}{45\tau}+\frac{5}{21\tau}\frac{\pi_\parallel}{\epsilon}-\frac{5}{21\tau}\frac{\pi_{\indep}}{\epsilon} -\frac{\pi_\parallel}{\epsilon^2}\left( \frac{\pi_\parallel + \pi_{\indep}}{2\tau}\right),\label{par}\\
      \frac{d}{d\tau}\left(\frac{\pi_{\indep}}{\epsilon}\right) + \Sigma\frac{\pi_{\indep}}{\epsilon} - \frac{2qB}{5T} \frac{\delta\hat{\pi}_{\indep}}{\epsilon}& = \frac{8}{15\tau}-\frac{5}{7\tau}\frac{\pi_\parallel}{\epsilon}-\frac{5}{21\tau}\frac{\pi_{\indep}}{\epsilon}-\frac{\pi_{\indep}}{\epsilon^2}\left( \frac{\pi_\parallel + \pi_{\indep}}{2\tau}\right),\label{perp}\\
      \frac{d}{d\tau}\left(\frac{\delta\hat{\pi}_{\indep}}{\epsilon}\right) + \Sigma^{'} \frac{\delta\hat{\pi}_{\indep}}{\epsilon} + \frac{2qB}{5T} \frac{\pi_{\indep}}{\epsilon}& = -\frac{5}{21\tau}\frac{\delta\hat{\pi}_{\indep}}{\epsilon}-\frac{\delta\hat{\pi}_{\indep}}{\epsilon^2}\left( \frac{\pi_\parallel + \pi_{\indep}}{2\tau}\right),\label{relperp}
 \end{align}
where, $\pi_{\indep} \equiv \pi^{-}_{\indep} + \ \pi^{+}_{\indep}$, $\delta\pi_{\indep} \equiv \delta\pi^{-}_{\indep} - \delta\pi^{+}_{\indep}$ and $\delta\pi_{\indep} = i\delta\hat{\pi}_{\indep}$. We have also considered $\Sigma^{'} $ = $4\Sigma/3$.
 Our objective is to solve Eqs.\ \eqref{energy}--\eqref{relperp} and investigate the behavior of the longitudinal and transverse components of the shear stress tensor in the presence of a magnetic field. Results are shown in Fig.\ 1.
\begin{figure}[H]
    \centering
    \subfigure[\ Longitudinal component]{\includegraphics[width=0.47\linewidth]{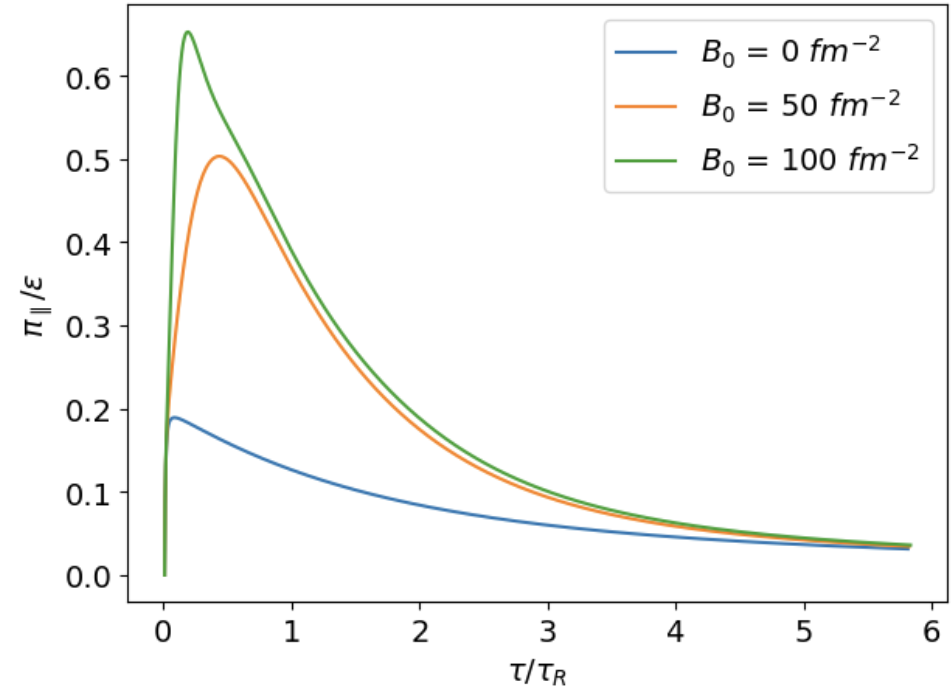}}
    \quad
    \subfigure[\ Transverse component ]{\includegraphics[width=0.47\linewidth]{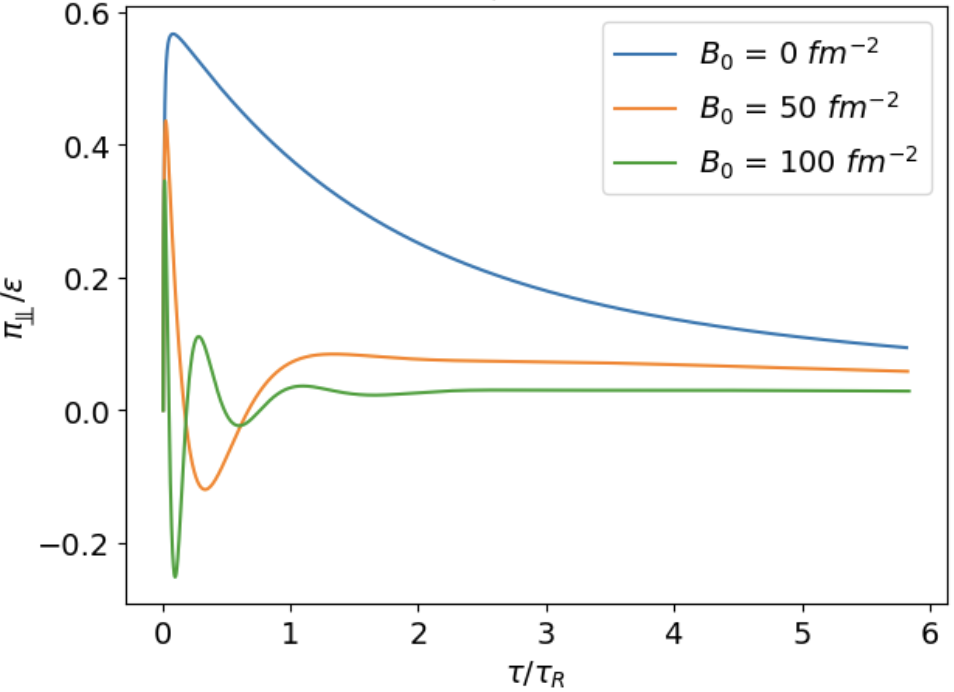}}
    \caption{Oscillatory dynamics in the transverse shear stress tensor are evident for higher initial magnetic field values ($\eta/s = 10$). A slight hint of oscillatory behavior is also observed in the longitudinal component with a stronger magnetic field.} 
    \label{fig: figbr2}
\end{figure}
An initial observation reveals distinct evolutionary paths for the different components of the shear stress tensor, each governed by its unique dynamical equation. Particularly noteworthy is the prominent oscillatory behavior observed in the transverse component, especially for higher values of $\eta/s$. While oscillations are present even for smaller $\eta/s$ values, they are less apparent in comparison. Such oscillatory behavior cannot be described by Israel-Stewart-like theories.

\section{Conclusion}
Hence, the presence of oscillatory dynamics is noted for moderately stronger magnetic fields. The Israel-Stewart theory functions effectively within a domain characterized by smaller magnetic fields, which have minimal impact on the system's evolution. However, for stronger magnetic fields, the theory proves inadequate, necessitating the examination of more fundamental equations to accurately capture the underlying dynamics. Our theoretical framework predicts the existence of oscillatory dynamics in the presence of higher magnetic field values, a scenario relevant to LHC and RHIC.

\bibliography{references}

\end{document}